%% file: paper.tex
\documentclass[conference,a4paper]{IEEEtran}
\pdfoutput=1

\usepackage[T1]{fontenc}
\usepackage[utf8]{inputenc}

\usepackage{amsmath,amssymb}
\usepackage{comment}
\usepackage{cite}
\usepackage{graphicx}
\usepackage{lipsum}
\usepackage{xcolor}
\usepackage{xurl}

\usepackage[acronym]{glossaries}
\glsdisablehyper
\newacronym{fspl}{FSPL}{free-space path loss}
\newacronym{isac}{ISAC}{integrated sensing and communication}
\newacronym{rcs}{RCS}{radar cross section}
\newacronym{rx}{Rx}{receiver}
\newacronym{snr}{SNR}{signal-to-noise ratio}
\newacronym{tx}{Tx}{transmitter}
\newacronym{vna}{VNA}{vector network analyzer}

\usepackage{siunitx}
\DeclareSIUnit{\dBc}{\deci\bel c}
\DeclareSIUnit{\dBm}{\deci\bel m}
\DeclareSIUnit{\dBFS}{\deci\bel FS}
\DeclareSIUnit{\dBi}{\deci\bel i}

\usepackage{hyperref}
\hypersetup{
	pdfauthor={Carsten Andrich, Isabella Varga, Tobias F. Nowack, Alexander Ihlow, Sebastian Giehl, Michael Schubert, Reiner S. Thomä, Matthias A. Hein},
	pdftitle={Wideband Antenna Deconvolution for Bistatic Millimeter Wave Radar Reflectivity Measurements},
    linkcolor={red!50!black},
    citecolor={blue!50!black},
    urlcolor={blue!80!black}
}

\usepackage{orcidlink}

\usepackage{tikz}
\usetikzlibrary{external}
\usetikzlibrary{positioning,calc}
\usetikzlibrary{arrows,shapes}

\usepackage{pgfplots}
\pgfplotsset{compat=1.18}
\pgfplotsset{plot coordinates/math parser=false}
\pgfplotsset{footnotesize}

\newcommand\encircle[1]{\raisebox{.5pt}{\textcircled{\raisebox{-.7pt}{\scalebox{0.9}{\strut #1}}}}}

\title{Wideband Antenna Deconvolution for Bistatic Millimeter Wave Radar Reflectivity Measurements}

\author{%
    \IEEEauthorblockN{%
        Carsten~Andrich\IEEEauthorrefmark{1}\IEEEauthorrefmark{2}\raisebox{.5ex}{\orcidlink{0000-0002-4795-3517}},
        Isabella~Varga\IEEEauthorrefmark{1}\IEEEauthorrefmark{3}\raisebox{.5ex}{\orcidlink{0009-0001-9391-8122}},
        Tobias~F.~Nowack\IEEEauthorrefmark{1}\raisebox{.5ex}{\orcidlink{0009-0003-5775-9851}},
        Alexander~Ihlow\IEEEauthorrefmark{1}\IEEEauthorrefmark{2}\raisebox{.5ex}{\orcidlink{0000-0002-9714-4881}},\\
        Sebastian~Giehl\IEEEauthorrefmark{1}\IEEEauthorrefmark{2}\raisebox{.5ex}{\orcidlink{0009-0008-1672-1351}},
        Michael~Schubert\IEEEauthorrefmark{1}\IEEEauthorrefmark{2}\raisebox{.5ex}{\orcidlink{0009-0003-4843-2507}},
        Reiner~S.~Thomä\IEEEauthorrefmark{1}\IEEEauthorrefmark{2}\raisebox{.5ex}{\orcidlink{0000-0002-9254-814X}},
        Matthias~A.~Hein\IEEEauthorrefmark{1}\IEEEauthorrefmark{3}\raisebox{.5ex}{\orcidlink{0000-0002-8751-7760}}
    }
    \IEEEauthorblockA{\IEEEauthorrefmark{1}Thuringian Center of Innovation in Mobility, Ilmenau, Germany}
    \IEEEauthorblockA{\IEEEauthorrefmark{2}Institute of Information Technology, Technische Universität Ilmenau, Ilmenau, Germany}
    \IEEEauthorblockA{\IEEEauthorrefmark{3}RF and Microwave Research Group, Technische Universität Ilmenau, Ilmenau, Germany}
}

\begin{document}

\maketitle

\begin{abstract}
Bistatic radar measurements offer unique spatial diversity and enhanced target characterization capabilities, rendering them increasingly vital for contemporary sensing application research.
The reliability of such measurements is contingent upon precise system and antenna calibration.
The prevailing technique is the substitution method, which involves the use of known reference objects.
We propose an over-the-air calibration algorithm for spherical bistatic measurement systems.
Our method is both significantly simpler and twice as fast as existing algorithms.
The application of our technique to reflectivity measurements of a metal sphere from 76 to 81 GHz demonstrates a dynamic range enhancement of up to 40 dB when compared with uncalibrated data.
A comparison with simulation data demonstrates a high degree of agreement between measurement and simulation.
\end{abstract}


\section{Introduction}
Fundamental research on bistatic wave propagation and radar reflectivity is highly relevant for environmental sensing, e.g. for automotive, aerospace and defense applications, as well as the upcoming mobile communication standard 6G, which includes \gls{isac} features that rely on bistatic geometries~\cite{liu24_book_isac,shatov24_access_jrc}.
As the first mobile communication standard, 6G is foreseen to employ instantaneous bandwidths in excess of several hundred MHz that enable radar applications with range resolutions in the order of decimeters or better.
This improvement facilitates micro-Doppler analysis~\cite{chen19_book_udoppler_radar} by spatially resolving individual target components, e.g., drone propellers~\cite{costa25_jsteap_model}, vehicle tires, or individual limbs~\cite{arxiv24andrich_bira}, to aid target or gesture identification.

Time-domain radar reflectivity analysis has significantly higher calibration requirements than conventional \gls{rcs} calibration.
The \gls{rcs} describes the scalar, real-valued, frequency-variant, electromagnetically effective size of a static point-target with respect to backscattered power.
As a generalization, the complex-valued radar reflectivity is capable of representing the impulse response of spatially extended targets with non-zero delay spread.
Thus, radar reflectivity calibration must compensate for time-domain effects that would otherwise convolve into and distort the measured target response.
These effects include the impulse responses of the measurement system and antennas in their installed state as well as parasitic reflections from lab installations like mechanical positioners.

This paper is structured as follows:
First, \autoref{sec:sota} briefly summarizes the current state of the art for \gls{rcs} calibration and \autoref{sec:algorithm} explains our proposed new algorithm.
Then, \autoref{sec:measurement_setup} outlines the measurement setup used to demonstrate the algorithm and \autoref{sec:results} discusses the measurement results.
Finally, \autoref{sec:conclusions} concludes this paper.

\section{State of the Art}
\label{sec:sota}

System parameters such as antenna coupling, bistatic angle dependence, and geometric alignment have a pronounced influence on measurement results. 
Various calibration approaches have been adopted to address these challenges. One method is to utilize the radar equation and calculate an angle-dependent calibration factor accounting for the measurement distances and antenna gains~\cite[p.\,75]{Diss_Schwind}. 
However, this requires precise knowledge of the measurement distances, antenna gains and any losses within the system.

The most common method employs reference targets with analytically known radar cross sections, such as metallic spheres, dihedral reflectors, or trihedral corner reflectors~\cite{IEEE_Std_1502}.
Among these, the metallic sphere remains widely used, especially for bistatic calibration approaches, due to its pronounced scattering over a wide angular range~\cite{balanis}.
The definition of different calibration types is motivated by the varying requirements and practical limitations encountered in radar measurement setups. Depending on the measurement objective and the desired accuracy, it is not always feasible or necessary to perform a full absolute calibration of the entire system. Radar cross section calibration methods are typically classified into three different calibration types according to the extent of the correction applied \cite{calibration_bradley}. The Type-1 calibration corrects amplitude and phase, whereby a complex-valued constant scale factor is applied. It is implemented as
\begin{equation}
    \sigma_\text{Target,cal} = \frac{\vert S_\text{21,Target} \vert ^2}{\vert S_\text{21,Ref} \vert ^2} \cdot \sigma_\text{Ref,theoretical},
\end{equation}
where $\sigma_\text{Target,cal}$ denotes the calibrated response of the target, $\vert S_\text{21,Target} \vert ^2$ and $\vert S_\text{21,Ref} \vert ^2$ the measured transmission data of the target and the reference object, respectively, and $\sigma_\text{Ref,theoretical}$ the theoretical RCS of the calibration object. The simple polarimetric calibration is known as Type-2 and compensates for the effects of zero-order antenna polarization distortions. The co-polarized subsystem distortion terms $R_\text{HH}, T_\text{HH}, R_\text{VV}, T_\text{VV}$ cannot be identified individually, but are determined as combinations of the receiver ($R_\text{??}$) and transmitter ($T_\text{??}$) antenna distortion factors. The correction matrix is
\begin{equation}
C = \begin{bmatrix} R_\text{HH}T_\text{HH} & 0 & 0 & 0 \\ 0 & R_\text{VV}T_\text{HH}  & 0 & 0 \\ 0 & 0 & R_\text{HH}T_\text{VV} \\ 0 & 0 & 0 & R_\text{VV}T_\text{VV}\end{bmatrix}.
\end{equation}
The calibrated scattering matrix $S$ is determined by
\begin{equation}
    S = C^{-1} (M - B)
\end{equation}
with $M$ as the measured scattering matrix and $B$ as the scattering matrix of the background.
Two calibration objects are needed, first a reference object with a known theoretical solution, same as for Type-1 calibration, and secondly a depolarizing object with a known scattering matrix for cross-polarization correction. Type-3 calibration needs three known calibration objects for a fully polarimetric correction and results in the correction matrix
\begin{equation}
C = \begin{bmatrix} 
R_\text{HH}T_\text{HH} & R_\text{HV}T_\text{HH} & R_\text{HH}T_\text{HV} & R_\text{HV}T_\text{HV} \\ 
R_\text{VH}T_\text{HH} & R_\text{VV}T_\text{HH} & R_\text{VH}T_\text{HV} & R_\text{VV}T_\text{HV} \\
R_\text{HH}T_\text{VH} & R_\text{HV}T_\text{VH} & R_\text{HH}T_\text{VV} & R_\text{HV}T_\text{VV} \\
R_\text{VH}T_\text{VH} & R_\text{VV}T_\text{VH} & R_\text{VH}T_\text{VV} & R_\text{VV}T_\text{VV}
\end{bmatrix}.
\end{equation}
The difference between this and Type-2 is that the coefficients of the distortion matrices are solved for separately \cite{Diss_Schwind, calibration_bradley}. 

Type-1 to Type-3 calibration necessitates supplementary reference measurements for all combinations of azimuth and elevation angles measured with the target, resulting in at least a duplication of the overall effort and measurement duration associated with this technique.

\section{Calibration Algorithm}
\label{sec:algorithm}

We propose a straightforward bistatic radar reflectivity calibration algorithm for spherical positioning systems.
Traditional Type-1 to~3 calibration requires one or more additional full measurement cycles with known reference objects.
In contrast, our algorithm does not necessitate a full measurement cycle of all bistatic angle permutations.
It relies on a single S\textsubscript{21}-parameter measurement with anti-parallel \gls{tx} and \gls{rx} antenna main lobe directions as depicted in \autoref{fig:bistatic_geometry}.
Consequently, the algorithm is only applicable to spherical bistatic systems capable of this geometry.
Equivalent to the state-of-the-art substitution method, our proposed algorithm requires far-field conditions, which depend primarily on target size and measurement frequency.
However, the consideration of near-field conditions is beyond the scope of this paper.

\subsection{Theory}

\begin{figure}
	\centering
	\input{Figures/geometry.tex}
    \caption{
        Simplified spherical bistatic measurement geometry with in-plane bistatic angle~$\beta$.
        \Gls{tx} and \gls{rx} antenna aperture planes move tangentially on two concentric spheres (here: circles) with the radii $R_\text{tx}$ and $R_\text{rx}$.
        Diagram combines both calibration $S_\text{cal}$ (\textcolor{blue}{blue}) and radar measurement $S_\text{radar}$ (\textcolor{red}{red}).
        The target is not present for the calibration measurement.
        Note that the spherical geometry ensures that both antennas face the center for all bistatic angles~$\beta$.
    }
    \label{fig:bistatic_geometry}
\end{figure}
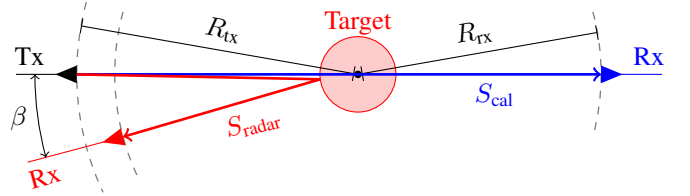

The calibration measurement S\textsubscript{21}-parameter $S_\text{cal}(f)$ is modeled by Eq.~(\ref{eq:fspl}), a slightly extended version of the well-known \gls{fspl} equation:
\begin{equation}\label{eq:fspl}
    |S_\text{cal}|^2
    = \frac{P_\text{rx}}{P_\text{tx}}
    = \frac{G_\text{sys} \cdot G_\text{tx} \cdot G_\text{rx} \cdot \lambda^2}{(4\pi)^2 \cdot (R_\text{tx}+R_\text{rx})^2}
\end{equation}
The \gls{tx} antenna gain $G_\text{tx}(f,\varphi,\theta)$ and \gls{rx} antenna gain $G_\text{rx}(f,\varphi,\theta)$ depend on frequency $f$, azimuth angle $\varphi$, and elevation angle $\theta$.
Additionally, $G_\text{sys}(f)$ represents the gain of the entire measurement system between both antenna connectors.
It includes the frequency response of cables, amplifiers, filters, frequency converters, and the used measurement device, e.g., a \gls{vna}.
Similarly, for bistatic radar reflectivity measurements $S_\text{radar}(f)$, the radar equation multiplied by $G_\text{sys}(f)$ applies:
\begin{equation}\label{eq:radar}
    |S_\text{radar}|^2
    = \frac{P_\text{rx}}{P_\text{tx}}
    = \frac{G_\text{sys} \cdot G_\text{tx} \cdot G_\text{rx} \cdot \lambda^2 \cdot \sigma}
           {(4\pi)^3 \cdot R_\text{tx}^2 \cdot R_\text{rx}^2}
\end{equation}

Now, divide the measured, complex-valued radar reflectivity $S_\text{radar}(f)$ by the \gls{fspl} calibration measurement $S_\text{cal}(f)$, thus deconvolving the antenna and system responses from the radar measurement:
\begin{equation}
\begin{aligned}
    \frac{|S_\text{radar}|^2}{|S_\text{cal}|^2} & =
        \frac{G_\text{sys} \cdot G_\text{tx} \cdot G_\text{rx} \cdot \lambda^2 \cdot \sigma}
             {(4\pi)^3 \cdot R_\text{tx}^2 \cdot R_\text{rx}^2}
    \cdot \frac{(4\pi)^2 \cdot (R_\text{tx}+R_\text{rx})^2}
               {G_\text{sys} \cdot G_\text{tx} \cdot G_\text{rx} \cdot \lambda^2} \\
    & = \frac{\sigma \cdot (R_\text{tx}+R_\text{rx})^2}
             {4\pi \cdot R_\text{tx}^2 \cdot R_\text{rx}^2}
\end{aligned}
\end{equation}
The radiation patterns of the antennas in their installed state cancel out for spherical measurement systems that ensure both antennas always face the target (cf. \autoref{fig:bistatic_geometry}).
Additionally, the result is now easily solved for the reflectivity
\begin{equation}
    \sigma = \left| \frac{S_\text{radar}}{S_\text{cal}} \right|^2 \cdot \frac{4\pi \cdot R_\text{tx}^2 \cdot R_\text{rx}^2}{(R_\text{tx}+R_\text{rx})^2},
\end{equation}
which enables scaling the calibrated measurement values to effective cross sections.
For identical aperture plane radii $R = R_\text{tx} = R_\text{rx}$ the equation simplifies further to
\begin{equation}
    \sigma = \left| \frac{S_\text{radar}}{S_\text{cal}} \right|^2 \cdot \pi R^2.
\end{equation}

\subsection{Application}
Apply the algorithm as follows:
Measure the \gls{fspl} between \gls{rx} and \gls{tx} antennas directly opposing each other.
Apply time gating to restrict the calibration measurement to within the measurement volume centered around the focal point of the spherical positioning system (cf. \autoref{fig:bistatic_geometry}).
While not strictly necessary, time gating will significantly reduce the noise incurred in the next step.
Time gating algorithms that operate directly in the frequency-domain~\cite{gedschold23_tap_uwb_radar_targets} will exhibit less truncation error than gating algorithms that operate by transforming from frequency-domain to time-domain and back.
Then, measure the radar reflectivity of the target.
Optionally, conduct a background measurement and apply background subtraction~\cite{IEEE_Std_1502}.
Finally, divide the resulting complex-valued frequency response by the time-gated calibration response and apply another time gating as required.
As a noteworthy side-effect, this deconvolution implicitly unwinds the entire propagation delay, resulting in the focal point being situated at the origin of any calibrated impulse response.

\section{Measurement Setup}
\label{sec:measurement_setup}

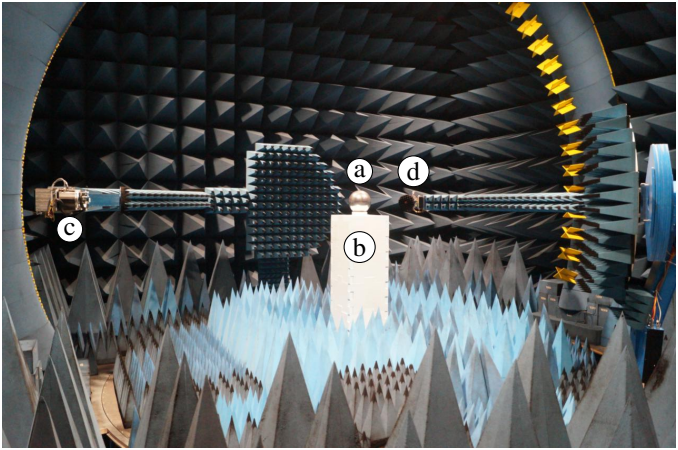
\begin{figure}[t!]
    \centering
    \begin{tikzpicture}
    \node[anchor=south west,inner sep=0] (image) at (0,0) {\includegraphics[width=\linewidth]{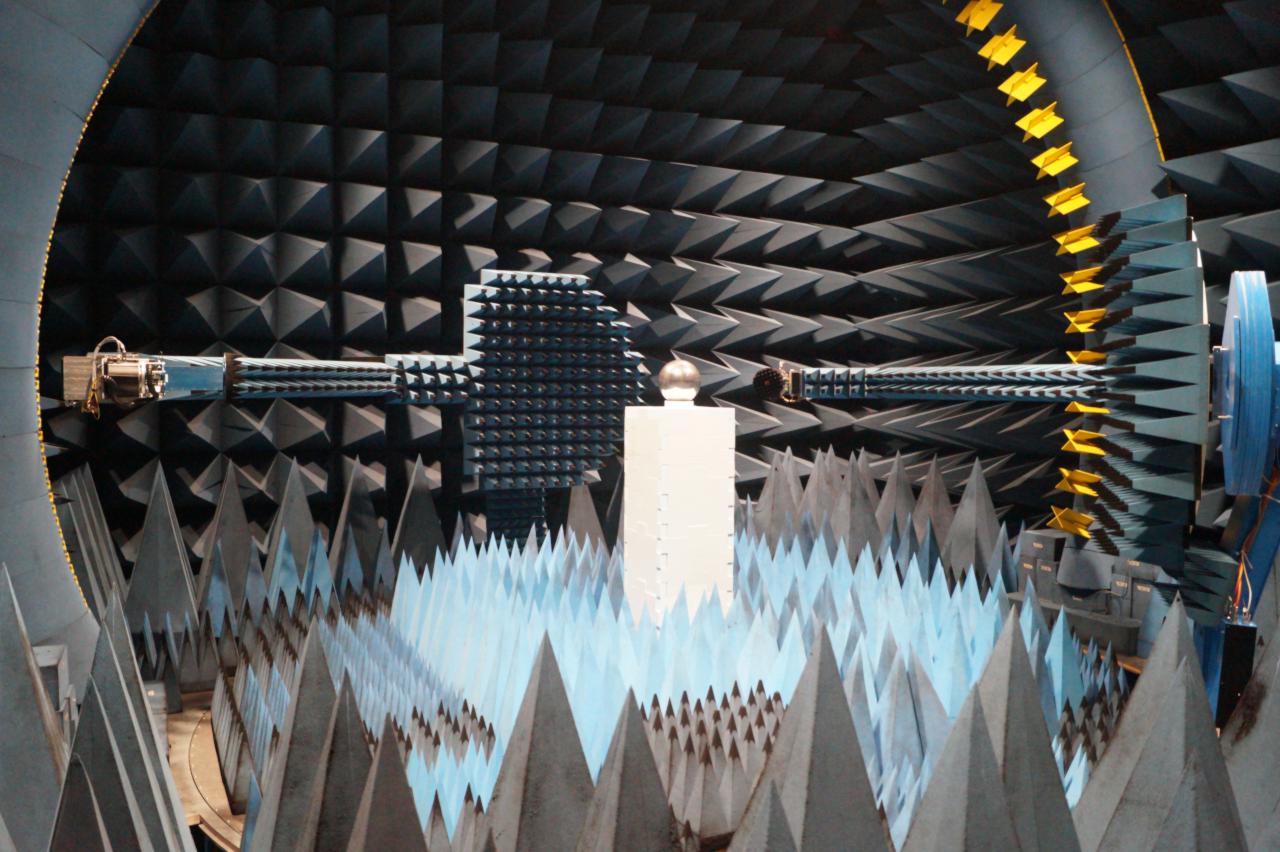}};
    \begin{scope}[x={(image.south east)},y={(image.north west)}]
        \node[minimum width=1ex,inner sep=.25ex,circle,fill=white,draw] () at (0.61,0.62) {d};
        \node[minimum width=1ex,inner sep=.25ex,circle,fill=white,draw] () at (0.10,0.49) {c};
        \node[minimum width=1ex,inner sep=.25ex,circle,fill=white,draw] () at (0.53,0.45) {b};
        \node[minimum width=1ex,inner sep=.25ex,circle,fill=white,draw] () at (0.53,0.62) {a};
    \end{scope}
    \end{tikzpicture}
    \caption{
        The Bistatic Radar (BIRA) measurement facility at the Thuringian Center of Innovation in Mobility~\cite{arxiv24andrich_bira, hein23_roe_vista}.
        The target, a metallic sphere~\encircle{a} with diameter 30\,cm, is positioned in the center on a styrodur pillar~\encircle{b}.
        The receiver~\encircle{c} remained stationary and the transmitter~\encircle{d} moved to realize a horizontal azimuth cut with both antennas fixed at co-elevation $\theta = 90^\circ$.
    }
    \label{fig:bira_sphere_wr12}
\end{figure}

\begin{table}[b]
    \caption{Overview of RF measurement parameters}
    \small
    \centering
        \begin{tabular}{ll}
        \hline
        Measurement parameter                                       & Value and unit                                           \\  \hline
        Frequency range $f$                                         & 76 \dots (0.005) \dots  81 GHz                   \\
        Antenna polarization                                        & VV and HH                                       \\
        Typical gain                                                & 25 dBi                                          \\
        3\,dB beamwidth (E-Plane)                                   & $9^\circ$                                       \\
        3\,dB beamwidth (H-Plane)                                   & $10^\circ$                                      \\
        Distance antennas to center                                 & 3.04\,m                                         \\
        Bi-static angle $\beta$                                     & $8^\circ \dots (1^\circ) \dots 243^\circ$       \\
        Tx, Rx co-elevation $\theta_\text{obs}, \theta_\text{ill}$     & $90^\circ$                                       \\ \hline
        \end{tabular}
    \label{tab:RFParam}
\end{table}

The authors operate the Bistatic Radar (BIRA) measurement facility depicted in \autoref{fig:bira_sphere_wr12} at the Thuringian Center of Innovation in Mobility at the Technische Universität Ilmenau, Ilmenau, Germany~\cite{arxiv24andrich_bira, hein23_roe_vista}.
BIRA comprises two independent spherical positioners:
The transmitter can move in azimuth and elevation and the receiver is fixed in azimuth and movable in elevation around the target positioned in the center.
For the purpose of this paper, measurements were conducted with the full azimuthal range of the transmitter resulting in bistatic angles of $\beta = 8^\circ \dots 243^\circ$ and both gantry arms positioned at the co-elevation of $\theta_\text{ill} = \theta_\text{obs} = 90^\circ$.

The bistatic measurements were performed using a Keysight N5222B \gls{vna} synchronized with an external Keysight N5173B signal generator.
Both supply the IF and LO signals, resp., for feeding external Virginia Diodes WR12CCU-B-M4 and WR12CCD-B-M4 millimeter wave converter modules.
The resulting frequency range is 76...81\,GHz, sampled at 1001 frequency points.
Two MI-WAVE 261E-25/387 pyramidal horn antennas were used and measurements were performed vertically (VV) and horizontally (HH) co-polarized.
The target, an easily reproducible metallic sphere of diameter $d$ = 30\,cm, was positioned on top of a styrodur pillar, to ensure that the antennas and the sphere shared the same height.
The measurement parameters are summarized in \autoref{tab:RFParam}.

\section{Results}
\label{sec:results}

\begin{figure}[t!]
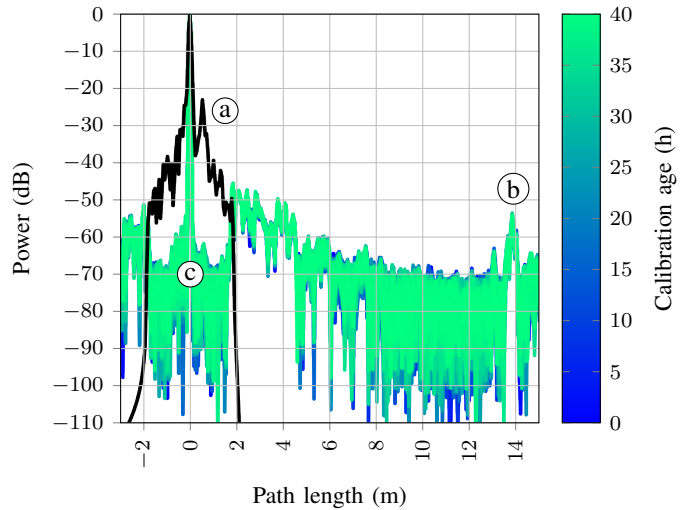

    \centering
    \include{Figures/meas_cal/stability_scatter}
    \caption{
        Power delay profile (PDP) of antenna calibration measurement without a target.
        Line-of-sight between Tx and Rx normalized to 0\,m and 0\,dB.
        Raw measurement (black) exhibits shallow roll-off and a discrete parasitic reflection~\encircle{a} off the positioners.
        In contrast, the deconvolved PDP (blue to green) has sharp edges and up to 40\,dB better dynamic range within the sweet spot~\encircle{c}.
        Note that the dynamic range remains stable over time.
        A double reflection~\encircle{b} between the Tx and Rx positioners is also visible.
    }
    \label{fig:calibration_time}
\end{figure}

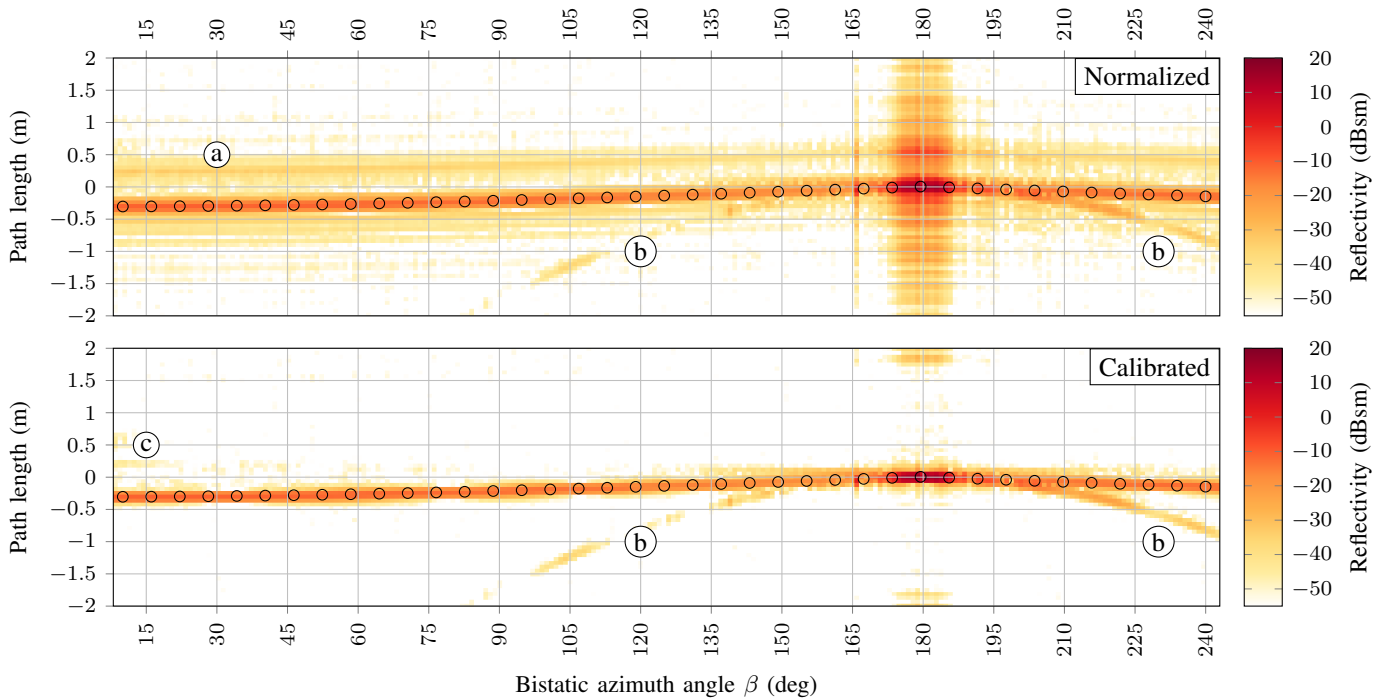
\begin{figure*}
    \centering
    \input{Figures/meas_sphere/heatmap}
    \caption{
        Bistatic radar reflectivity of a metal sphere with diameter 30\,cm measured from 76 to 81\,GHz in the horizontal plane with vertical polarization.
        Top plot with normalization for comparability only.
        Note the parasitic peak~\encircle{a}, which corresponds to \autoref{fig:calibration_time}\,\encircle{a}.
        Bottom plot with antenna calibration and deconvolution.
        Note the significant improvement due to the deconvolution.
        The imperfect background subtraction leaves residual antenna crosstalk~\encircle{b} and styrodur pillar response~\encircle{c}.
        The strong path represents the tangential reflection off the sphere or the diffraction around the sphere in case of forward scattering, respectively.
        The measured bistatic range coincides with a geometric path model (circles) of the described propagation effects.
    }
    \label{fig:sphere_impresp}
\end{figure*}

\subsection{Calibration Measurement}

\autoref{fig:calibration_time} illustrates the results of the calibration measurement as a power delay profile (PDP).
The PDP of the measurement antennas in their installed state is clearly affected by the mechanical positioners, see the parasitic reflection in \autoref{fig:calibration_time}~\encircle{a}.
We opted for a $\pm2$\,m time gating window size to exemplify both target response and antenna crosstalk while also removing most clutter, e.\,g., from the antenna arch surrounding the positioners.
Within the window, our calibration algorithm improves the PDP substantially, sharpening it into a distinct peak with minimal delay spread.
The calibrated response has sharp edges and a dynamic range improved by up to 40\,dB.

To assess the stability of the calibration, we carried out a continuous 40\,h measurement.
In \autoref{fig:calibration_time}, the first response serves as the calibration value for all subsequent measurements to visualize the stability over calibration age as a trace color transition from blue to green.
Over the entire measurement duration, the calibration remained highly stable with no measurable loss of dynamic range.

\subsection{Sphere}

After the calibration, we installed the pillar and sphere (cf. \autoref{fig:bira_sphere_wr12}) and measured an azimuth cut in the horizontal plane.
Measurement results with and without the calibration are shown in \autoref{fig:sphere_impresp}.
The uncalibrated results are normalized in terms of amplitude and delay so that results are directly comparable.
Note the significant improvement visible as a much sharper PDP, similar to the effect displayed in \autoref{fig:calibration_time}.
We rely on a geometric model of the tangential reflection off the sphere to analytically describe the expected PDP.
Within the calibrated measurement results, the only visible artifacts are from residual echos off the styrodur pillar and the crosstalk between \gls{tx} and \gls{rx}.
Both leak through the background subtraction due to imperfect coherence between the background and the sphere measurements.

\subsection{Simulation/Verification}

\begin{figure*}
    \centering
    \input{Figures/sim_sphere/heatmap}
    \caption{
        Comparison of measured (same as \autoref{fig:sphere_impresp}) and simulated radar reflectivity of sphere.
        Bottom plot shows absolute reflectivity difference (measurement minus simulation) for values above --55\,dBsm.
        Note that the reflection off the sphere (marked by circles) shows excellent agreement within 3\,dB except for forward scattering ($\beta \approx 150^\circ \dots\, 210^\circ$).
        Antenna crosstalk, i.e., line of sight~\encircle{b}, and the styrodur sphere mount~\encircle{c} are clearly visible.
    }
    \label{fig:sphere_impresp_sim}
\end{figure*}
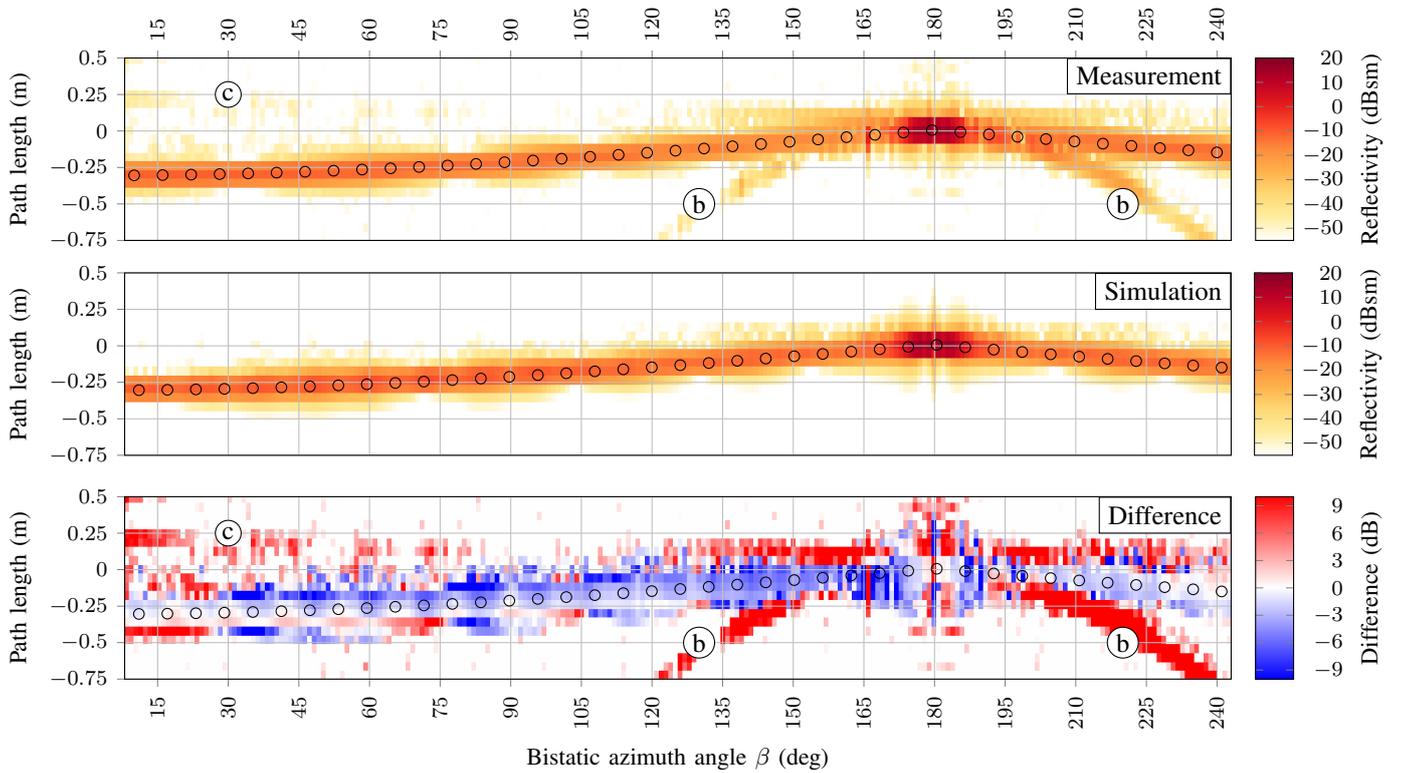

To verify the measurement results shown in \autoref{fig:sphere_impresp}, we also simulated in the Ansys Electronics Desktop HFSS numerical simulation environment using its SBR+ solver.
The geometry and simulation parameters resemble the measurement setup as described in \autoref{sec:measurement_setup}.
\autoref{fig:sphere_impresp_sim} compares the measurement and the simulation results.
Most notable deviations are caused by the styrodur pillar, which was not simulated, and the antenna crosstalk, which the simulated background subtraction removes fully due to perfect coherence.
Otherwise, the measurement and simulation values match very well.

\section{Conclusions}
\label{sec:conclusions}

We motivated, derived, and verified a novel calibration algorithm for spherical bistatic radar reflectivity measurement by means of measurement and simulation.
The calibration algorithm improved the dynamic range of the measurements by up to 40\,dB, also enabling good agreement between measurement and simulation data.
Compared to the state-of-the-art Type-1 to~3 \gls{rcs} calibration algorithms (cf. \autoref{sec:sota}), our calibration method provides significant improvement:
It does not require a reference object, obviating the intricacies of mounting and aligning it, and it needs only a single S\textsubscript{21}-measurement, i.e., one \gls{vna} sweep, instead of extra time-consuming, exhaustive measurement cycles of all bistatic angle constellations used for the target object.
This results in an acceleration by at least factor two compared to a conventional Type-1 calibration.

Both state-of-the-art \gls{rcs} calibration techniques and our proposed method are -- theoretically -- only applicable in the far-field and with geometrically small targets.
In practice, we have found our algorithm to deliver reliable results even with spatially extended targets~\cite[Fig.\,15]{arxiv24andrich_bira}.
Possible future improvements include the integration of synthetic aperture techniques to fully support calibration under near-field conditions and with large targets.

\section*{Acknowledgments}
The research was funded by the Federal State of Thuringia, Germany, and the European Social Fund (ESF) under grants 2017 FGI 0007 (project ``BiRa''), 2021 FGI 0007 (project ``Kreatör''), and 2023 IZN 0005 (project ``research initiative digital mobility'').
The research was also funded by the German Federal Ministry of Research, Technology and Space under grant 16KIS2225 (project ``6Gsens'').

\bibliographystyle{IEEEtran}
\bibliography{paper}

\end{document}

%% file: Figures/geometry.tex
\begin{tikzpicture}

\def\Rtx{3.2}

\node at (0,0) [draw, red, fill=red!20, circle, minimum size=10mm] (Targ) {};
\node at (0,0) [fill=black, circle, inner sep=1pt]{};
\node at (Targ.north) [red, anchor=south, yshift=-1mm] {Target};

\draw [line width=0.5pt, dashed, gray] (165:3.7) arc (165:205:3.7);
\draw [line width=0.5pt, dashed, gray] (165:3.2) arc (165:205:3.2);
\draw [line width=0.5pt, dashed, gray] (15:3.2) arc (15:-15:3.2);
\draw [|-|, black] (170:3.7) -- (170:1.85) node[anchor=south, rotate=-10] {$R_\text{tx}$} -- (170:0.05);
\draw [|-|, black] (10:3.2) -- (10:1.6) node[anchor=south, rotate=10] {$R_\text{rx}$} -- (10:0.05);

\node at (180:4) [isosceles triangle, fill=black, inner xsep=0mm, anchor=apex, rotate=180] (Tx) {};
\node at (Tx.apex) [anchor=south east] {Tx};
\draw (Tx.center) -- (180:4.5);
\node at (0:3.5) [isosceles triangle, fill=blue, inner xsep=0mm, anchor=apex, rotate=0] (Rxc) {};
\node at (Rxc.apex) [blue, anchor=south west] {Rx};
\draw[blue] (Rxc.center) -- (0:4.0);
\node at (195:3.5) [isosceles triangle, fill=red, inner xsep=0mm, anchor=apex, rotate=195] (Rxm) {};
\node at (195:4.0) [red, anchor=north east, rotate=15] {Rx};
\draw[red] (Rxm.center) -- (195:4.5);
\node at (187.5:4.5) [black] {$\beta$};
\draw [<->, black] (180:4.25) arc (180:195:4.25);

\draw [->, blue, line width=1pt, solid] (180:3.7) -- (0:3.2);
\node at (0:1.8) [blue, anchor=north] {$S_\text{cal}$};
\draw [->, red, line width=1pt, solid, line join=bevel] (180:3.7) -- (187.5:0.5) -- (195:3.2);
\node at (195:1.5) [red, anchor=north, rotate=15] {$S_\text{radar}$};
\end{tikzpicture}

%% file: Figures/meas_cal/stability_scatter.tex
\begin{tikzpicture}
\pgfdeclarelayer{foreground}
\pgfsetlayers{main,foreground}
\begin{axis}[
        domain=8:243,
        colorbar,
        colormap={winter}{
            rgb=(0.0000, 0.0000, 1.0000),
            rgb=(0.0000, 0.1255, 0.9373),
            rgb=(0.0000, 0.2510, 0.8745),
            rgb=(0.0000, 0.3765, 0.8118),
            rgb=(0.0000, 0.5020, 0.7490),
            rgb=(0.0000, 0.6275, 0.6863),
            rgb=(0.0000, 0.7529, 0.6235),
            rgb=(0.0000, 0.8784, 0.5608),
            rgb=(0.0000, 1.0000, 0.5000),
        },
        colorbar style={
            ylabel={Calibration age (h)},
            ytick={0,5,...,40},
            yticklabel style={
                text width=1em,
                align=left
            }
        },
        width=0.8\linewidth,
        height=7cm,
        enlargelimits=false,
        axis on top,
        tick align=outside,
        tick pos=left,
        xtick pos=bottom,
        xlabel={Normalized bistatic range (m)},
        ylabel={Normalized power (dB)},
        xmajorgrids,
        ymajorgrids,
        xtick distance=2,
        ytick distance=10,
        x tick label style={rotate=90,anchor=east}
    ]
    \addplot[point meta min=0, point meta max=40] graphics[xmin=-3, xmax=15, ymin=-110, ymax=0] {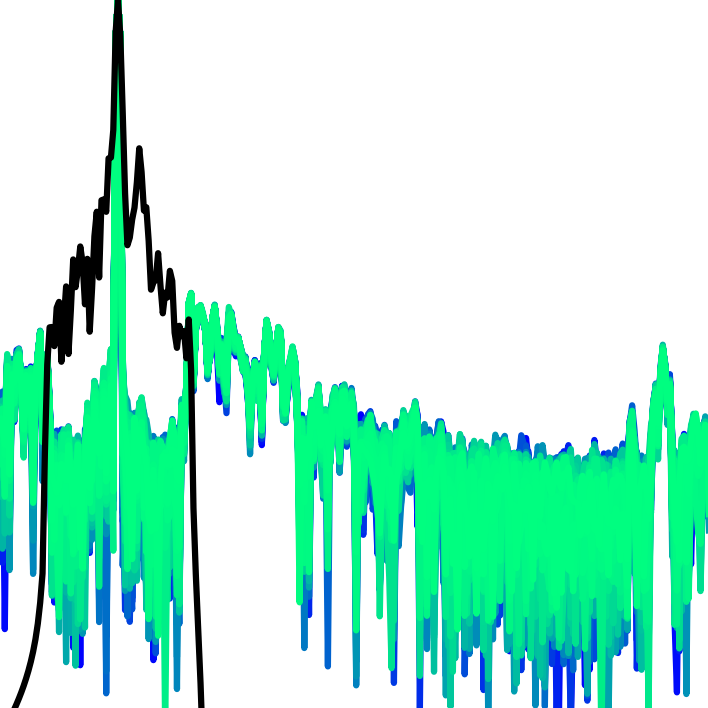};
    \begin{pgfonlayer}{foreground}
        \node[minimum width=1ex,inner sep=.25ex,circle,fill=white,draw] () at (1.5,-26) {a};
        \node[minimum width=1ex,inner sep=.25ex,circle,fill=white,draw] () at (13.9,-47) {b};
        \node[minimum width=1ex,inner sep=.25ex,circle,fill=white,draw] () at (0,-70) {c};
    \end{pgfonlayer}
\end{axis}
\end{tikzpicture}

%% file: Figures/meas_sphere/heatmap.tex
\begin{tikzpicture}
\pgfdeclarelayer{foreground}
\pgfsetlayers{main,foreground}
\begin{axis}[
        domain=8:243,
        colorbar,
        colormap={WhYlOrRd}{
            rgb=(1.0000, 1.0000, 1.0000),
            rgb=(1.0000, 0.9291, 0.6268),
            rgb=(0.9961, 0.8498, 0.4615),
            rgb=(0.9960, 0.6963, 0.2973),
            rgb=(0.9921, 0.5491, 0.2342),
            rgb=(0.9863, 0.3019, 0.1636),
            rgb=(0.8867, 0.0996, 0.1107),
            rgb=(0.7346, 0.0000, 0.1490),
            rgb=(0.5020, 0.0000, 0.1490)
        },
        colorbar style={
            ylabel={Reflectivity (dBsm)},
            ytick={-50,-40,...,20},
            yticklabel style={
                text width=1.5em,
                align=right
            }
        },
        width=0.89\linewidth,
        height=5cm,
        enlargelimits=false,
        axis on top,
        tick align=outside,
        tick pos=left,
        xtick pos=top,
        title={Normalized},
        every axis title/.style={below left,at={(1,1)},draw=black,fill=white},
        ylabel={Bistatic range (m)},
        xmajorgrids,
        ymajorgrids,
        xtick distance=15,
        ytick distance=0.5,
        x tick label style={rotate=90,anchor=west}
    ]
    \addplot[point meta min=-55, point meta max=20] graphics[xmin=8, xmax=243, ymin=-2, ymax=2] {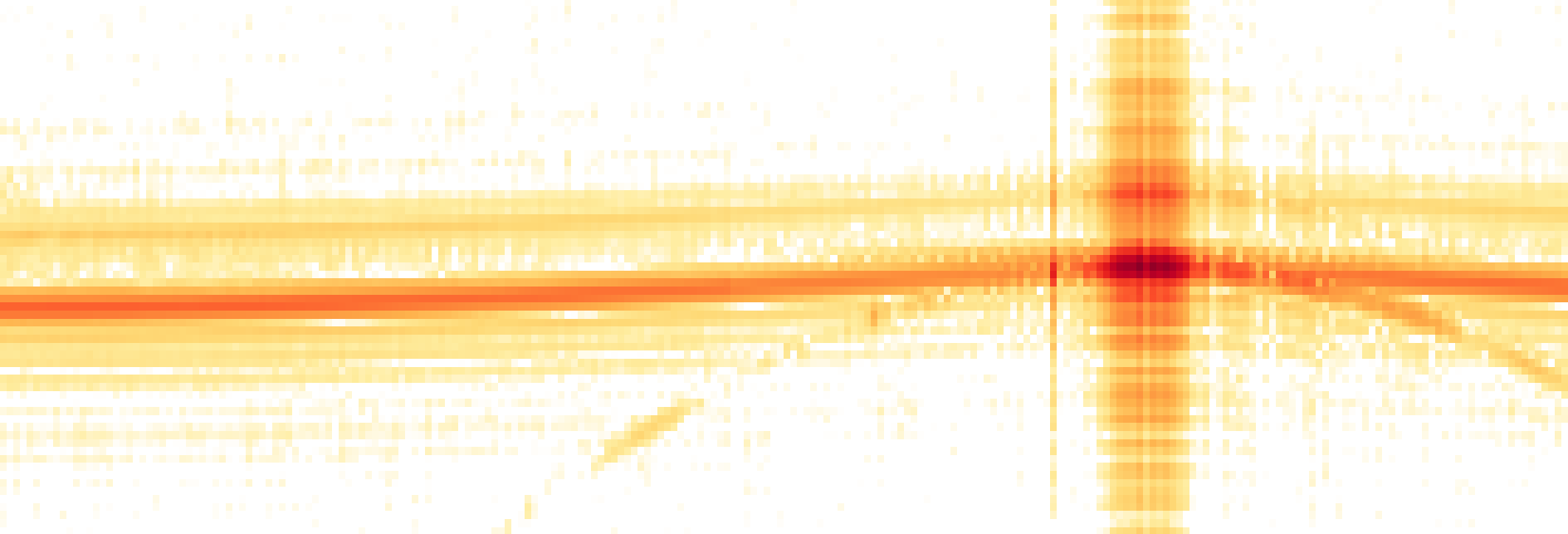};
    \addplot[mark=o, mark repeat = 6, mark phase = 3,samples=234,only marks] {2*((3.04 - 0.3048/2*abs(cos(x/2)))^2 + (0.3048/2*abs(sin(x/2)))^2)^0.5 - 2*3.04};
    \begin{pgfonlayer}{foreground}
        \node[minimum width=1ex,inner sep=.25ex,circle,fill=white,draw] () at (30,0.5) {a};
        \node[minimum width=1ex,inner sep=.25ex,circle,fill=white,draw] () at (120,-1) {b};
        \node[minimum width=1ex,inner sep=.25ex,circle,fill=white,draw] () at (230,-1) {b};
    \end{pgfonlayer}
\end{axis}
\end{tikzpicture}

\begin{tikzpicture}
\pgfdeclarelayer{foreground}
\pgfsetlayers{main,foreground}
\begin{axis}[
        domain=8:243,
        colorbar,
        colormap={WhYlOrRd}{
            rgb=(1.0000, 1.0000, 1.0000),
            rgb=(1.0000, 0.9291, 0.6268),
            rgb=(0.9961, 0.8498, 0.4615),
            rgb=(0.9960, 0.6963, 0.2973),
            rgb=(0.9921, 0.5491, 0.2342),
            rgb=(0.9863, 0.3019, 0.1636),
            rgb=(0.8867, 0.0996, 0.1107),
            rgb=(0.7346, 0.0000, 0.1490),
            rgb=(0.5020, 0.0000, 0.1490)
        },
        colorbar style={
            ylabel={Reflectivity (dBsm)},
            ytick={-50,-40,...,20},
            yticklabel style={
                text width=1.5em,
                align=right
            }
        },
        width=0.89\linewidth,
        height=5cm,
        enlargelimits=false,
        axis on top,
        tick align=outside,
        tick pos=left,
        xtick pos=bottom,
        title={Calibrated},
        every axis title/.style={below left,at={(1,1)},draw=black,fill=white},
        xlabel={Bistatic azimuth angle $\beta$ (deg)},
        ylabel={Bistatic range (m)},
        xmajorgrids,
        ymajorgrids,
        xtick distance=15,
        ytick distance=0.5,
        x tick label style={rotate=90,anchor=east}
    ]
    \addplot[point meta min=-55, point meta max=20] graphics[xmin=8, xmax=243, ymin=-2, ymax=2] {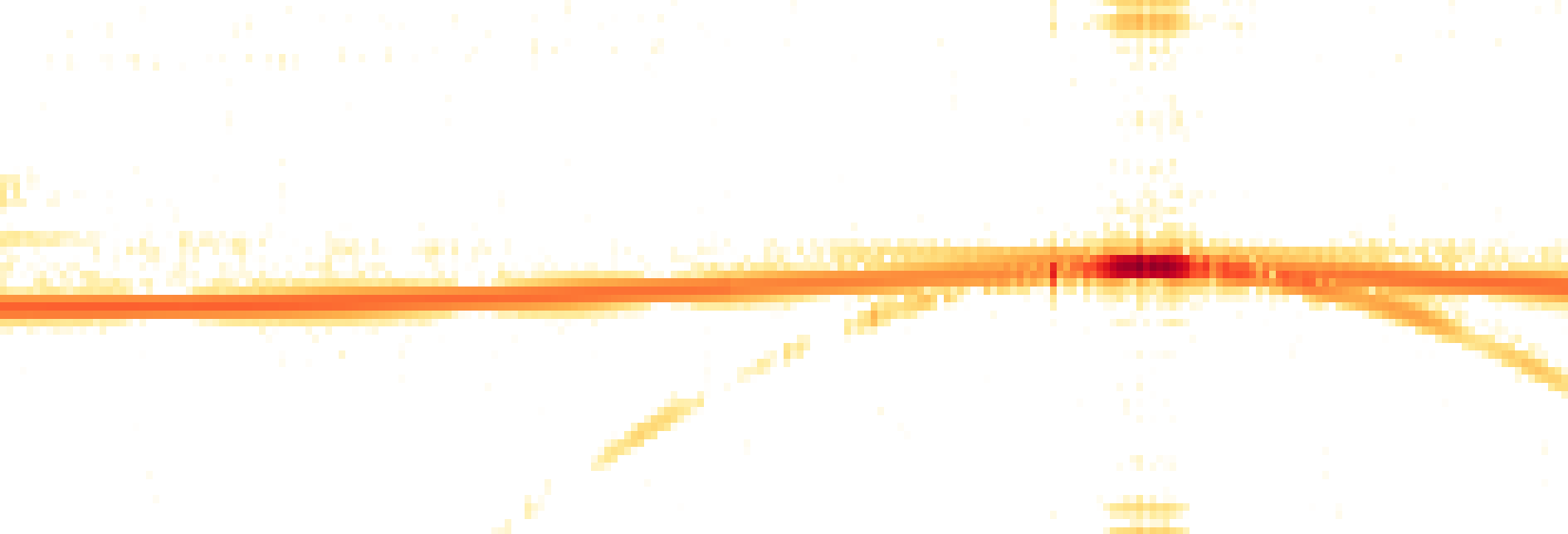};
    \addplot[mark=o, mark repeat = 6, mark phase = 3,samples=234,only marks] {2*((3.04 - 0.3048/2*abs(cos(x/2)))^2 + (0.3048/2*abs(sin(x/2)))^2)^0.5 - 2*3.04};
    \begin{pgfonlayer}{foreground}
        \node[minimum width=1ex,inner sep=.25ex,circle,fill=white,draw] () at (120,-1) {b};
        \node[minimum width=1ex,inner sep=.25ex,circle,fill=white,draw] () at (230,-1) {b};
        \node[minimum width=1ex,inner sep=.25ex,circle,fill=white,draw] () at (15,0.5) {c};
    \end{pgfonlayer}
\end{axis}
\end{tikzpicture}

%% file: Figures/sim_sphere/heatmap.tex
\begin{tikzpicture}
\pgfdeclarelayer{foreground}
\pgfsetlayers{main,foreground}
\begin{axis}[
        domain=8:243,
        colorbar,
        colormap={WhYlOrRd}{
            rgb=(1.0000, 1.0000, 1.0000),
            rgb=(1.0000, 0.9291, 0.6268),
            rgb=(0.9961, 0.8498, 0.4615),
            rgb=(0.9960, 0.6963, 0.2973),
            rgb=(0.9921, 0.5491, 0.2342),
            rgb=(0.9863, 0.3019, 0.1636),
            rgb=(0.8867, 0.0996, 0.1107),
            rgb=(0.7346, 0.0000, 0.1490),
            rgb=(0.5020, 0.0000, 0.1490)
        },
        colorbar style={
            ylabel={Reflectivity (dBsm)},
            ytick={-50,-40,...,20},
            yticklabel style={
                text width=1.5em,
                align=right
            }
        },
        width=0.89\linewidth,
        height=4cm,
        enlargelimits=false,
        axis on top,
        tick align=outside,
        tick pos=left,
        xtick pos=top,
        title={Measurement},
        every axis title/.style={below left,at={(1,1)},draw=black,fill=white},
        ylabel={Bistatic range (m)},
        xmajorgrids,
        ymajorgrids,
        xtick distance=15,
        ytick distance=0.25,
        x tick label style={rotate=90,anchor=west}
    ]
    \addplot[point meta min=-55, point meta max=20] graphics[xmin=8, xmax=243, ymin=-0.75, ymax=0.5] {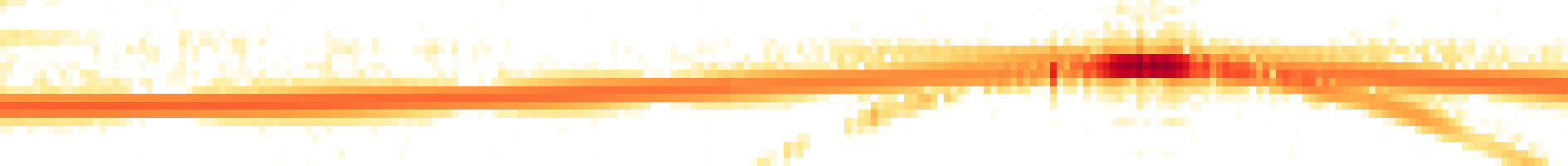};
    \addplot[mark=o, mark repeat = 6, mark phase = 3,samples=234,only marks] {2*((3.04 - 0.3048/2*abs(cos(x/2)))^2 + (0.3048/2*abs(sin(x/2)))^2)^0.5 - 2*3.04};
    \begin{pgfonlayer}{foreground}
        \node[minimum width=1ex,inner sep=.25ex,circle,fill=white,draw] () at (130,-0.5) {b};
        \node[minimum width=1ex,inner sep=.25ex,circle,fill=white,draw] () at (220,-0.5) {b};
        \node[minimum width=1ex,inner sep=.25ex,circle,fill=white,draw] () at (30,0.25) {c};
    \end{pgfonlayer}
\end{axis}
\end{tikzpicture}

\begin{tikzpicture}
\pgfdeclarelayer{foreground}
\pgfsetlayers{main,foreground}
\begin{axis}[
        domain=8:243,
        colorbar,
        colormap={WhYlOrRd}{
            rgb=(1.0000, 1.0000, 1.0000),
            rgb=(1.0000, 0.9291, 0.6268),
            rgb=(0.9961, 0.8498, 0.4615),
            rgb=(0.9960, 0.6963, 0.2973),
            rgb=(0.9921, 0.5491, 0.2342),
            rgb=(0.9863, 0.3019, 0.1636),
            rgb=(0.8867, 0.0996, 0.1107),
            rgb=(0.7346, 0.0000, 0.1490),
            rgb=(0.5020, 0.0000, 0.1490)
        },
        colorbar style={
            ylabel={Reflectivity (dBsm)},
            ytick={-50,-40,...,20},
            yticklabel style={
                text width=1.5em,
                align=right
            }
        },
        width=0.89\linewidth,
        height=4cm,
        enlargelimits=false,
        axis on top,
        tick align=outside,
        tick pos=left,
        xtick style={draw=none},
        xtick pos=bottom,
        title={Simulation},
        every axis title/.style={below left,at={(1,1)},draw=black,fill=white},
        ylabel={Bistatic range (m)},
        xmajorgrids,
        ymajorgrids,
        xtick distance=15,
        ytick distance=0.25,
        xticklabel=\empty
    ]
    \addplot[point meta min=-55, point meta max=20] graphics[xmin=8, xmax=243, ymin=-0.75, ymax=0.5] {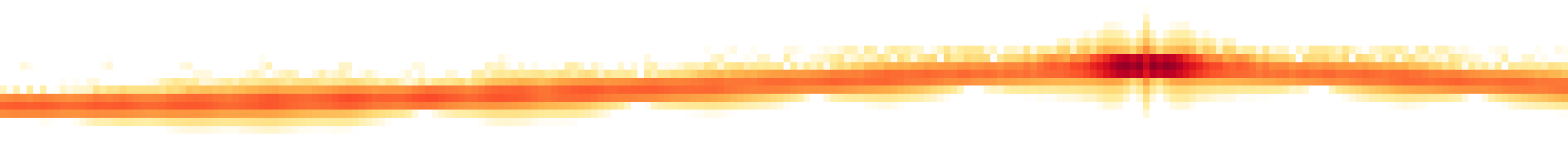};
    \addplot[mark=o, mark repeat = 6, mark phase = 4,samples=234,only marks] {2*((3.04 - 0.3048/2*abs(cos(x/2)))^2 + (0.3048/2*abs(sin(x/2)))^2)^0.5 - 2*3.04};
\end{axis}
\end{tikzpicture}

\begin{tikzpicture}
\pgfdeclarelayer{foreground}
\pgfsetlayers{main,foreground}
\begin{axis}[
        domain=8:243,
        colorbar,
        colormap={WhYlOrRd}{
            rgb=(0.0000, 0.0000, 1.0000),
            rgb=(0.2510, 0.2510, 1.0000),
            rgb=(0.5020, 0.5020, 1.0000),
            rgb=(0.7529, 0.7529, 1.0000),
            rgb=(1.0000, 0.9961, 0.9961),
            rgb=(1.0000, 0.7451, 0.7451),
            rgb=(1.0000, 0.4941, 0.4941),
            rgb=(1.0000, 0.2431, 0.2431),
            rgb=(1.0000, 0.0000, 0.0000),
        },
        colorbar style={
            ylabel={Difference (dB)},
            ytick={-9,-6,-3,0,3,6,9},
            yticklabel style={
                text width=1.5em,
                align=right
            }
        },
        width=0.89\linewidth,
        height=4cm,
        enlargelimits=false,
        axis on top,
        tick align=outside,
        tick pos=left,
        xtick pos=bottom,
        title={Difference},
        every axis title/.style={below left,at={(1,1)},draw=black,fill=white},
        xlabel={Bistatic azimuth angle $\beta$ (deg)},
        ylabel={Bistatic range (m)},
        xmajorgrids,
        ymajorgrids,
        xtick distance=15,
        ytick distance=0.25,
        x tick label style={rotate=90,anchor=east}
    ]
    \addplot[point meta min=-10, point meta max=10] graphics[xmin=8, xmax=243, ymin=-0.75, ymax=0.5] {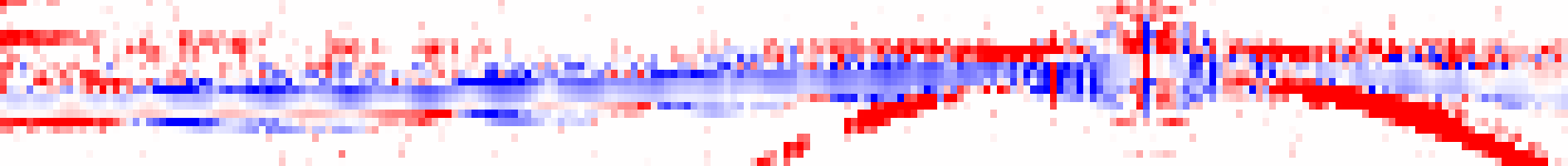};
    \addplot[mark=o, mark repeat = 6, mark phase = 4,samples=234,only marks] {2*((3.04 - 0.3048/2*abs(cos(x/2)))^2 + (0.3048/2*abs(sin(x/2)))^2)^0.5 - 2*3.04};
    \begin{pgfonlayer}{foreground}
        \node[minimum width=1ex,inner sep=.25ex,circle,fill=white,draw] () at (130,-0.5) {b};
        \node[minimum width=1ex,inner sep=.25ex,circle,fill=white,draw] () at (220,-0.5) {b};
        \node[minimum width=1ex,inner sep=.25ex,circle,fill=white,draw] () at (30,0.25) {c};
    \end{pgfonlayer}
\end{axis}
\end{tikzpicture}